# Induced Hilbert Space, Markov Chain, Diffusion Map and Fock Space in Thermophysics

Xing M. (Sherman) Wang
Sherman Visual Lab, Sunnyvale, CA, USA

## Table of Contents

## Abstract

In this article, we continue to explore *Probability Bracket Notation (PBN)*, proposed in our previous article. Using both Dirac vector bracket notation (*VBN*) and *PBN*, we define *induced Hilbert space* and *induced sample space*, and propose that there exists *an equivalence relation* between a Hilbert space and a sample space constructed from the same base observable(s). Then we investigate *Markov transition matrices* and their eigenvectors to make *diffusion map*s with two examples: a simple *graph theory example*, to serve as a prototype of bidirectional transition operator; a famous *text document example* in IR literature, to serve as a *tutorial* of diffusion map in text document space. We show that the sample space of the Markov chain and the Hilbert space spanned by the eigenvectors of the transition matrix are *not equivalent*. At the end, we apply our *PBN* and equivalence proposal to Thermophysics by associating sample (phase) space with the Hilbert space of a single particle and the Fock space of many-particle systems.

## 1. Introduction

In our previous article [1], we proposed a new set of symbols, the *Probability Bracket Notation (PBN)*. We demonstrated that *PBN* could play a similar role for probability theories [2] as Dirac Vector Bracket Notation (*VBN*) plays for Hilbert vector space.

In this article, we take advantage of the abstracting power of both *PBN* and *VBN* to explore the relations between Hilbert space and sample space. We introduce *base observables* (*operators)* for such spaces, and define the *induced sample spa*ce from a Hilbert space and the *induced Hilbert space* from a sample space. We propose that there exists *equivalence* between a Hilbert space and a sample space, if the two spaces are constructed on the same base observable(s). This provides us with a freedom of mapping system states from one space to another, and enables us to apply different ways to handle their *probability distribution functions* (PDF).

As a case study of *PBN*, *VBN* and the *equivalence* proposition, we investigate transition operator of Markov chains [2-3] in diffusion maps [4-6] for data clustering. We use two examples to show the procedure. First is a simple example from graph theory [7], which has a symmetric transition matrix, so can be used to build a *bidirectional* transition operator. The second example is from the famous 3-text-document collection [9-10]. This numerical example can serve as *a tutorial* of how to make diffusion map for text documents, and also provides a concrete evidence for the necessary condition of *equivalence* between a Hilbert space and a sample space.

Finally, we apply the equivalence relation between phase space in Thermophysics and the Hilbert space of a single particle or many-particle systems. We derive the wave function of a single particle of monatomic ideal gas in a square well at temperature $T$, and verify the consistence of statistic formulas in *PBN* and in the Fock space of identical bosons or fermions.

## 2. Induced Sample Space and Induced Hilbert Space

Dirac *VBN* has proved to be a very powerful tool to deal with states in Hilbert space. *PBN* might offer a similarly powerful tool to deal with states in sample space. Together, they provide a unified platform to investigate the relationship between the two spaces and their states.

In this section, with the help of both *PBN* and *VBN*, we show that how to build induced space from each other. Starting from a Hilbert space and its base observable, we can construct an *induced sample space*, and there is a one-to-one equivalence between a system state in the Hilbert and a corresponding state in the induced sample space. We can also construct an induced Hilbert space from a sample space and its base observable, and for each sample system state, there exists an equivalent Hilbert state. The equivalence is critical for mapping states between Hilbert space, sample space and the space of probability vectors.

## 2.1. From Hilbert Space to Induced Sample Space

**Definition 2.1.1** (*Base Observables of Hilbert Space*): If a Hilbert space $\mathcal{B}$ is spanned by eigenvectors of a set of Hermitian operators $\{\hat{H}\}$, then the elements of $\{\hat{H}\}$ are called the *base observables* or *base operators* of the Hilbert space $\mathcal{B}$.

**Definition 2.1.2** (*Base Observables of Sample Space*): If a sample space $\Omega$ is based on the outcomes of observing a set of random variables $\{\hat{X}\}$, the elements of $\{\hat{X}\}$ are called the *base observables* or *base operators* of the sample space.

Let us consider a Hilbert space in QM. We can build a v-basis from the complete set of normalized eigenvectors of a Hermitian operator (e.g., the Hamiltonian $\hat{H}$ of the system):

$$\hat{H}\,|\psi_i\rangle = E_n\,|\psi_i\rangle, \quad \sum_{i=1}^{n} |\psi_i\rangle\langle\psi_i| = I, \quad \langle\psi_i|\psi_j\rangle = \delta_{ij} \qquad (2.1.1)$$

These eigenvectors form a vector-basis. The state of the system now can be expanded as:

$$|\Psi\rangle = I\,|\Psi\rangle = \sum_{i=1}^{n} |\psi_i\rangle\langle\psi_i|\Psi\rangle = \sum_{i=1}^{n} c_i\,|\psi_i\rangle \qquad (2.1.2)$$

Its normalization requires:

$$\langle\Psi|\Psi\rangle = \sum_{i,j}^{n} \langle\psi_i|c_i{}^* c_j|\psi_j\rangle = \sum_i |c_i|^2 = 1 \qquad (2.1.3)$$

It is well known in QM that the expectation value of $\hat{H}$ is:

$$\begin{aligned}\langle\hat{H}\rangle &= \langle\Psi|\hat{H}|\Psi\rangle = \sum_{i,j}^{n} \langle\psi_i|c_i{}^* c_j \hat{H}|\psi_j\rangle \\ &= \sum_{i,j} \delta_{ij} E_j\,|c_j|^2 = \sum_j E_j\,|c_j|^2\end{aligned} \qquad (2.1.4)$$

Hence, the probability of the system found in $i$-th state is determined by a Hilbert PDF:

$$\{p_i:\ p_i \equiv |c_i|^2 = |\langle\psi_i|\Psi\rangle|^2\} \qquad (2.1.5)$$

By our definition, operator $\hat{H}$ is the base observable of the Hilbert space, with its v-basis defined in (2.1.1). Each PDF represents one of its possible system states.

Now let us do one-to-one map from the v-basis to *P*-basis of a sample space *Ω*, which is a CMD (complete mutually disjoint) set:

$$\sum_{i=1}^{n} | \psi_i ) P( \psi_i | = I, \quad P( \psi_i | \psi_j ) = \delta_{ij} \tag{2.1.6a}$$

The *base observable* $\hat{E}$ has its eigen-kets and eigen-bra as:

$$\hat{E} = \sum_{i=1}^{n} | \psi_i ) \, E_i \, P( \psi_i |, \quad \hat{E} \, | \psi_i ) = E_i \, | \psi_i ) \tag{2.1.6b}$$

Then, applying Eq. (2.2.4) of Ref. [1], the expectation value of *E* in the sample space is given by:

$$\langle E \rangle = P(\Omega | \hat{E} | \Omega) = \sum_{\psi_i \in \Omega} P(\Omega | \hat{E} | \psi_i) P(\psi_i | \Omega) = \sum_{\psi_i \in \Omega} P(\Omega | E_i | \psi_i) P(\psi_i | \Omega)$$

$$= \sum_{\psi_i \in \Omega} E_i \, P(\Omega | \psi_i) P(\psi_i | \Omega) = \sum_{\psi_i \in \Omega} E_i \, P(\psi_i | \Omega) \equiv \sum_i E_i m(i) \tag{2.1.7}$$

If the system state |Ω) correctly represents the Hilbert state in (2.1.2), then we should set the PDF and system state according to the following relation:

$$| \Omega ) = \sum | \psi_i ) P( \psi_i | \Omega ) = \sum m(i) \, | \psi_i ) = \sum | c_i |^2 \, | \psi_i ) \tag{2.1.8}$$

The system *P*-ket |Ω) belongs to the sample space, induced by the Hilbert space.

**Definition 2.1.3** (*Induced Sample Space*): A sample space Ω is induced from a Hilbert space *Ɓ*, if the basis and base observable set of Ω are constructed from the v-basis and base observable set of *Ɓ*.

**Definition 2.1.4** (*Equivalence of a Hilbert State and a Sample State*): A Hilbert state |Ψ⟩ and a sample state |Ω) are equivalent if and only if they have the same PDF with respect to the basis associated with the same base observable(s), i.e. they have the following isomorphism:

$$m(i) \equiv P(\psi_i | \Omega) \Leftrightarrow | \langle \psi_i | \Psi \rangle |^2 = | c_i |^2 \tag{2.1.9}$$

**Definition 2.1.5** (*Equivalence of a Hilbert Space and a Sample Space*): A Hilbert space and a sample space are equivalent if and only if the two spaces are constructed on the same base observable(s), i.e. they have the following isomorphism:

$$\sum_{i=1}^{n} | \psi_i ) P( \psi_i | = I, \quad P( \psi_i | \psi_j ) = \delta_{ij}, \quad P( \psi_i | \hat{X} | \psi_j ) = x_i \, \delta_{ij} \tag{2.1.10a}$$

$$\Leftrightarrow \sum_{i=1}^{n} |\psi_i\rangle\langle\psi_i| = I, \quad \langle\psi_i|\psi_j\rangle = \delta_{ij}, \quad \langle\psi_i|\hat{x}|\psi_j\rangle = x_i\delta_{ij} \tag{2.1.10b}$$

Now, for each system state or PDF in Hilbert space, we can find an equivalent sample state or PDF in the sample space. Reversely, for each system state or PDF in the sample space, we can find an equivalent physics state in Hilbert space:

$$|\Psi\rangle = \sum_{i=1}^{n} c_i |\psi_i\rangle, \quad |c_i| = \sqrt{m(i)} \tag{2.1.11}$$

Although $c_i$ is not unique in (2.1.11), physically, they represent the same state, because in Hilbert FDF only $|c_i|$ matters. This leads to our first proposition:

**Proposition 2.1.1:** Starting from a base observable in Hilbert space, we can build an *induced sample space*. The Hilbert space and the induced sample space are equivalent. There is bijective equivalence between a physical state in the Hilbert space and a system state in the induced sample space.

Using the Hilbert v-basis, as we did in §4.1 of Ref. [1], we can map the state $P$-ket (2.1.8) in the induced sample space as a probability column vector (PCV):

$$\langle i|\Omega\rangle = m(i) = |c_i|^2, \quad i \in \{1,2,...n\}, \quad |\Omega\rangle = \sum_1^n |i\rangle\langle i|\Omega\rangle = \sum_1^n m(i)|i\rangle \tag{2.1.12}$$

## 2.2. From Sample Space to Induced Hilbert Space

Now let us go back to our example 2.1.1 of Ref. [1] (rolling a die). We have a base observable in the **die sample space**, as described in Eqs. (2.1.19) of Ref [1],

$$\hat{X}|i) = i|i), \quad P(i|\hat{X} = P(i|i, \quad i = \{1,2,\ldots 6\} \tag{2.2.1}$$

The sample space has the following basis:

$$P(i|j) = \delta_{ij}, \quad \sum_{i=1}^{6} |i)P(i| = I. \tag{2.2.2}$$

Assuming uniform PDF, we obtain the only system state $P$-ket:

$$P(i|\Omega) = 1/6, \quad i \in \{1,2,...6\}, \quad |\Omega) = \sum_1^6 |i)P(i|\Omega) = \sum_1^6 \frac{1}{6}|i) \tag{2.2.3}$$

The expectation value of our observable, according to Eq. (2.2.4) of [1] is given by:

$$P(\Omega\,|\,\hat{X}\,|\,\Omega)=\sum_{i=1}^{6}P(\Omega\,|\,\hat{X}\,|\,i)P(i\,|\Omega)=\sum_{i=1}^{6}i\,P(i\,|\Omega)=\sum_{x=1}^{6}i/6 \tag{2.2.4}$$

Now let us do one-to-one map from the *P*-basis to Hilbert v-basis, which is:

$$\langle i\,|\,j\rangle=\delta_{ij},\quad \sum_{i=1}^{6}|\,i\rangle\langle i\,|=I. \tag{2.2.5}$$

We can define a Hermitian operator using v-basis as follows:

$$\hat{x}\equiv\sum_{i=1}^{6}|\,i\rangle\, i\,\langle i\,| \tag{2.2.6a}$$

It is the base observable of a Hilbert space and it has following eigenvectors:

$$\hat{x}\,|\,i\rangle=i\,|\,i\rangle \qquad i\in\{1,2,...6\} \tag{2.2.6b}$$

Now any state of the Hilbert space can be expanded as:

$$|\,\Psi\rangle\;=I\,|\,\Psi\,\rangle=\sum_{i=1}^{6}|\,i\rangle\langle i\,|\,\Psi\,\rangle=\sum_{i=1}^{n}c_i\,|\,i\rangle \tag{2.2.7}$$

And the expectation value of our base observable is given by Eq. (2.1.4):

$$\langle X\rangle=\langle\psi\,|\,\hat{x}\,|\psi\rangle=\sum_{i,j}^{n}\langle i\,|\,c_i{}^{*}c_j\hat{x}\,|\,j\rangle=\sum_{i,j}\delta_{ij}j\,|\,c_i\,|^2=\sum_{i=1}^{6}i\,|\,c_i\,|^2 \tag{2.2.8}$$

**Definition 2.2.1** (*Induced Hilbert Space*): A Hilbert space $\mathcal{B}$ is induced from a sample space Ω, if the base and base observable set of $\mathcal{B}$ are constructed from the *P*-basis and base observable set of Ω.

Therefore, the Hilbert space we constructed is induced from the sample space (2.2.1).

Comparing (2.2.8) with (2.2.4), we see that if we want our Hilbert state correctly reflect our die sample state, we must restrict our Hilbert state to following one:

$$|\,\psi\rangle\;=\;\sum_{i=1}^{n}c_i\,|\,i\rangle,\quad |\,c_i\,|=\sqrt{P(i\,|\,\Omega)}=\sqrt{1/6} \tag{2.2.9}$$

But *the reversal is not true*. The following state is a valid normalized system state of the Hilbert space:

$$|\psi\rangle = \sqrt{1/3}\,|1\rangle + \sqrt{1/6}\,|2\rangle + \sqrt{1/3}\,|3\rangle + \sqrt{1/12}\,|4\rangle + \sqrt{1/12}\,|5\rangle \qquad (2.2.8)$$

But there is no equivalent state in the *sample space of a fair die*, which has only on state *P*-ket, given by Eq. (2.2.3). This leads to our second proposition:

**Proposition 2.2.1:** Starting from a base observable set in sample space, we can build an induced Hilbert space. The induced Hilbert space and the sample space are equivalent. For any possible PDF in the sample space, there is an equivalent physical state in the induced Hilbert space.

Using the induced Hilbert v-basis, as we did in §4.1 of Ref. [1], we can map the state *P*-ket (2.2.3) to a probability column vector (PCV):

$$\langle i\,|\,\Omega\rangle = 1/6, \quad i \in \{1,2,\ldots 6\}, \quad |\,\Omega\rangle = \sum_{1}^{6} |\,i\rangle\langle i\,|\,\Omega\rangle = \sum_{1}^{6} \frac{1}{6}\,|\,i\rangle \qquad (2.2.10)$$

**Note:** As we have discussed in Proposition 2.1.5 of Ref. [1], the state *P*-ket $|\Omega)$ presents a system state with its PDF, but the state *P*-bra $(\Omega|$ just represents the union of all possible events.

## 3. Markov Chains and Diffusion Maps

*Markov chains* [2-3] describe the time evolution of system states in a sample space. By nature, the sample space has a base observable set and the basis associated with the observable. The sample system states are represented by *probability vectors* ([2], §11.1). Each probability vector is a snapshot of the *probability distribution function* (PDF) at a give time. The Markov transition matrix transforms PDF of current state to the PDF of next time. If the transition matrix satisfies certain conditions, its eigenvectors form a complete basis of a Hilbert space, called the diffusion space [4-6]. Then one can map data points in original sample space onto the diffusion space, with reduced dimension based on the order of eigenvalues. In this section, we give two detailed numerical examples to illustrate the procedures.

### 3.1. Time-Dependent Probability Vectors and Markov Chains

We assume our sample space has the following base observable with discrete *P*-basis:

$$\hat{X}\,|\,j) = j\,|\,j), \quad P(i\,|\,j) = \delta_{ij}, \quad \sum_{i=1}^{r} |\,i)P(i\,| = I \qquad (3.1.1a)$$

Using the *P*-basis, we can build an induced Hilbert space as described in Eq. (2.2.5-6).

$$\hat{X}\,|\,j\rangle = j\,|\,j\rangle,\quad \langle i\,|\,j\rangle = \delta_{ij},\quad \sum_{i=1}^{r} |\,i\rangle\langle i\,| = I \tag{3.1.1b}$$

A general time-dependent sample state in Ω can be decomposed as:

$$|\,\Omega^{(t)}) = I\,|\,\Omega^{(t)}) = \sum_{i}^{r} |\,i\rangle\langle i_i\,|\,\Omega^{(t)}) = \sum_{i}^{r} m^{(t)}(i)\,|\,i) = \begin{pmatrix} m^{(t)}(1) \\ m^{(t)}(2) \\ \vdots \\ m^{(t)}(r) \end{pmatrix} \tag{3.1.2}$$

As we have discussed, we can build a time-dependent probability column vector (PCV) from the above state:

$$|\,\Omega^{(t)}\rangle = I\,|\,\Omega^{(t)}\rangle = \sum_{i}^{r} |\,i\rangle\langle i\,|\,\Omega^{(t)}\rangle = \sum_{i}^{r} m^{(t)}(i)\,|\,i_i\rangle = \begin{pmatrix} m^{(t)}(1) \\ m^{(t)}(2) \\ \vdots \\ m^{(t)}(r) \end{pmatrix} \tag{3.1.3}$$

Its counter-part, a row vector, is given by Eq. (2.1.14b) of Ref [1], which is not a probability row vector (PRV) and has no PDF:

$$\langle\Omega\,| = \sum_{i}^{r} \langle i\,| = [1,\quad 1,\quad \cdots,\quad 1] \tag{3.1.4}$$

We can obtain a PRV with time-dependent distribution function as the transpose of PCV in (3.1.3):

$$\langle\Omega^{(t)}\,| = \sum_{i}^{r} m^{(t)}(i)\langle i\,| = [m^{(t)}(1), m^{(t)}(2), \ldots, m^{(t)}(r)] \tag{3.1.5}$$

The *transition matrix element* $P_{ij}$ *of a Markov chain is defined as* [2, 3]:

$$P_{ij} \equiv P(X_{t+1} = j\,|\,X_t = i) \equiv (X_{t+1} = j\,|\,X_t = i) \tag{3.1.5a}$$

$$\sum_{j=1}^{r} P_{ij} = 1 \tag{3.1.5b}$$

In matrix form, if we define a PRV at t = 0 as $u^{(0)}$, then the left operating of $\boldsymbol{P}$ n times on it will give the PRV at time = $n$ ([2], theorem 11.2):

$$u^{(n)} = u^{(0)}P^n, \; or: \quad u^{(n)}{}_i = \sum_{j=1}^{r} u^{(n)}{}_j P^n{}_{ji} \tag{3.1.5c}$$

The *left-acting operator*, to act on a PRV from right, is defined as (see [1], Eq. (4.2.5)):

$$\overleftarrow{P} = \sum_{i',j'=1}^{r} |\,i'\rangle p_{i'j'} \langle j'|, \quad \langle \Omega^{(t)} \,|\, \overleftarrow{P} = \langle \Omega^{(t+1)} \,| \tag{3.1.6}$$

Our second example (§3.4) will use the right acting transition operator.

If we define a transition matrix as the transpose of (3.1.5a):

$$\vec{p}_{ji} \equiv p^T{}_{ji} = p_{ij} \tag{3.1.7a}$$

Then we can build a right-acting transition operator acting on a PCV, already given by Eq. (3.1.3).

$$u^{(n)} = \vec{P}^n u^{(0)}, \; or: \quad u^{(n)}{}_i = \sum_{j=1}^{r} \vec{P}^n{}_{ij} u^{(0)}{}_j \tag{3.1.7b}$$

The *right-acting operator*, to act on a PCV from left, can be defined as:

$$\vec{P} = \sum_{i',j'=1}^{r} |\,i'\rangle p_{j'i'} \langle j'|, \quad \vec{P} \,|\, \Omega^{(t)} \rangle = |\, \Omega^{(t+1)} \rangle \tag{3.1.8}$$

If the transition matrix is symmetric, $p_{ij} = p_{ji}$, then the corresponding operator becomes bidirectional. Our first example (§3.3) will use a symmetric transition matrix.

In summary, starting from the base observable and the basis of a sample space of a Markov chain, we can build an *induced Hilbert space*. The sample state *P*-kets can be mapped to probability vectors using the induced Hilbert base. The Markov transition matrix can be constructed as a *transition operator* (left-acting, right-acting or bidirectional) represented in the induced Hilbert base.

## 3.2. Markov chain and Diffusion Map

In this section, we introduce the definitions and notations about diffusion map in Ref. [4-6]. They are summarized as follows:

1. A *data set* is defined by $\Omega = \{x_1, x_2, \ldots, x_n\}$.
2. A *graph* is constructed as G = (Ω, W), where each x is a point, corresponding to a node of the graph and are connected by an edge with a *non-negative and symmetric weight:*

$$w(x,y)=w(y,x) \tag{3.2.1a}$$

3. *The degree* of a node by:

$$d(x)=\sum_{z\in X} w(x,z) \tag{3.2.1b}$$

4. The an $n$ by $n$ transformation matrix $\boldsymbol{P}$ defined by

$$p(x,y)=\frac{w(x,y)}{d(x)} \tag{3.2.2}$$

5. This matrix is a *transition matrix of a Markov chain*, because:

$$\sum_{y\in X} p(x,y)=1, \quad p(x,y)\geq 0 \tag{3.2.3a}$$

6. The transition is assumed to be *reversible*, i.e., there exists a row vector $\boldsymbol{\pi}$ such that:

$$\pi(x)p(x,y)=\pi(y)p(y,x) \tag{3.2.3b}$$

7. With the assumed properties, $P$ has a sequence of eigenvalues:

$$1=\lambda_0 >|\lambda_1|\geq \ldots \geq |\lambda_n|\geq 0 \tag{3.2.4a}$$

The corresponding right eigenvectors are (after t-step):

$$\boldsymbol{P}^t \psi_m = \lambda_m{}^t \psi_m \tag{3.2.4b}$$

8. The so-called diffusion coordinates are introduced by a *diffusion mapping*

$$\Psi_t : \{\Omega \mapsto \Re^{n-1}\} \tag{3.2.5a}$$

$$\Psi_t : x \mapsto \begin{pmatrix} \lambda_1{}^t \psi_1(x) \\ \lambda_2{}^t \psi_2(x) \\ \vdots \\ \lambda_{n-1}{}^t \psi_{n-1}(x) \end{pmatrix} \tag{3.2.5b}$$

The 0-th eigenvector is not included because it is a fixed vector with uniform distribution function.

9. The diffusion distance at time t is given by:

$$D_t^2(x,z) = \sum_{j=1}^{n-1} \lambda_j^{\;2t} (\psi_j(x) - \psi_j(z))^2 = \| \Psi_t(x) - \Psi_t(z) \|^2 \tag{3.2.6}$$

Because of the conditions of the matrix, it has a set of real eigenvalues as in (3.2.4) and a complete set of orthogonal left and right eigenvectors ([5], Appendix B):

$$\hat{P} | \psi_i \rangle = \lambda_i | \psi_i \rangle, \quad \sum_{i=1}^{n} | \psi_i \rangle \langle \varphi_i | = I, \quad \langle \varphi_i | \psi_j \rangle = \delta_{ij} \tag{3.2.7}$$

Now the diffusion map can be expressed as:

$$\Psi_t : x \mapsto \Psi_t(x) \equiv \langle x | \Psi_t \rangle = \begin{pmatrix} \lambda_1^{\;t} \langle x | \psi_1 \rangle \\ \lambda_2^{\;t} \langle x | \psi_2 \rangle \\ \vdots \\ \lambda_{n-1}^{\;t} \langle x | \psi_{n-1} \rangle \end{pmatrix} \tag{3.2.8}$$

The diffusion distance now is:

$$D_t^2(x,z) = \sum_{j=1}^{n-1} \lambda_j^{\;2t} (\langle x | \psi_j \rangle - \langle z | \psi_j \rangle)^2 = \| \langle x | \Psi_t \rangle - \langle z | \Psi_t \rangle \|^2 \tag{3.2.9}$$

### 3.3. Diffusion Map: a Graph Theory Example

As the first numerical example, let us use the following small graph example.

**Example 3.3.1 The Four-point Delta Graph** (a symmetric case, see Fig. 1b, [7]):

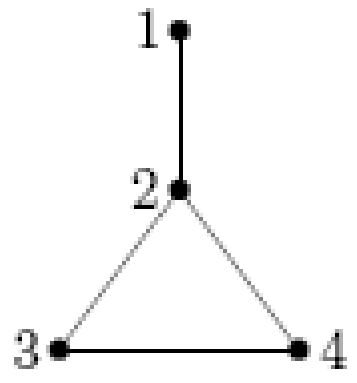

$$\vec{P} = \begin{bmatrix} 6/11 & 5/11 & 0 & 0 \\ 5/11 & 0 & 3/11 & 3/11 \\ 0 & 3/11 & 4/11 & 4/11 \\ 0 & 3/11 & 4/11 & 4/11 \end{bmatrix} \tag{3.3.1a}$$

We select this one because the transition matrix is symmetric, so we can have a bidirectional transition operator. The sample space of this graph has a basis as in Eq. (3.1.1) with $r = 4$. Its base observable is its edge number, with value from 1 to 4:

$$\hat{n} | j) = j | j), \quad P(i | \hat{n} | j) = j \delta_{ij}, \quad \sum_{i=1}^{4} | i) P(i | = 1 \tag{3.3.1b}$$

The sample state $P$-ket in this sample space is:

$$|\Omega^{(t)}) = \sum_i^4 |i)P(i|\Omega^{(t)}) = \sum_i^4 m^{(t)}(i)|i) \tag{3.3.1c}$$

The matrix given in Eq. (3.3.1a) satisfies all the conditions in Eq. (3.2.1-4.3.4) and more. We have four eigenvectors which are normalized as:

$$\hat{P}|\psi_i\rangle = \lambda_i |\psi_i\rangle, \quad \sum_{i=1}^4 |\psi_i\rangle\langle\psi_i| = I, \quad \langle\psi_i|\psi_j\rangle = \delta_{ij} \tag{3.3.1d}$$

Because the matrix of (3.3.1a) is symmetric, its left and right eigenvectors are simply the transpose of each other. The first one is the fixed PCV corresponding to eigenvalue = 1. As a Hilbert vector, its normalized form is:

$$\hat{P}|\psi_0\rangle = |\psi_0\rangle, \quad |\psi_0\rangle = \begin{pmatrix} 1/\sqrt{4} \\ 1/\sqrt{4} \\ 1/\sqrt{4} \\ 1/\sqrt{4} \end{pmatrix} \tag{3.3.2a}$$

The rest three eigenvector are neither PCV nor PRV.

$$\hat{P}|\psi_1\rangle = \frac{7}{11}|\psi_1\rangle, \quad \hat{P}|\psi_2\rangle = \frac{-4}{11}|\psi_2\rangle, \quad \hat{P}|\psi_3\rangle = 0\cdot|\psi_3\rangle \tag{3.3.2b}$$

$$|\psi_1\rangle = \frac{1}{2\sqrt{11}}\begin{pmatrix} -5 \\ -1 \\ 3 \\ 3 \end{pmatrix}, \quad |\psi_2\rangle = \frac{1}{\sqrt{22}}\begin{pmatrix} 2 \\ -4 \\ 1 \\ 1 \end{pmatrix}, \quad |\psi_3\rangle = \frac{1}{\sqrt{2}}\begin{pmatrix} 0 \\ 0 \\ -1 \\ 1 \end{pmatrix} \tag{3.3.2c}$$

Document $d_1$, $d_2$, $d_3$ and $d_4$ now are mapped to a 3-dimensional space:

$$\Psi_t : d_i \mapsto \Psi_t(i) \equiv \langle i|\Psi_t\rangle = \begin{pmatrix} \lambda_1{}^t \langle i|\psi_1\rangle \\ \lambda_2{}^t \langle i|\psi_2\rangle \\ \lambda_3{}^t \langle i|\psi_3\rangle \end{pmatrix} \tag{3.3.3}$$

Or, using the values we have:

$$d_1 \mapsto \Psi_t(1) = \begin{pmatrix} \left(\frac{7}{11}\right)^t \left(\frac{-5}{2\sqrt{11}}\right) \\ \left(\frac{-4}{11}\right)^t \left(\frac{2}{\sqrt{22}}\right) \\ \frac{0}{\sqrt{2}} \end{pmatrix}, \quad d_2 \mapsto \Psi_t(2) = \begin{pmatrix} \left(\frac{7}{11}\right)^t \left(\frac{-1}{2\sqrt{11}}\right) \\ \left(\frac{-4}{11}\right)^t \left(\frac{-4}{\sqrt{22}}\right) \\ \frac{0}{\sqrt{2}} \end{pmatrix} \tag{3.3.4a}$$

$$d_3 \mapsto \Psi_t(3) = \begin{pmatrix} \left(\frac{7}{11}\right)^t \left(\frac{3}{2\sqrt{11}}\right) \\ \left(\frac{-4}{11}\right)^t \left(\frac{1}{\sqrt{22}}\right) \\ \frac{0}{\sqrt{2}} \end{pmatrix}, \quad d_4 \mapsto \Psi_t(4) = \begin{pmatrix} \left(\frac{7}{11}\right)^t \left(\frac{3}{2\sqrt{11}}\right) \\ \left(\frac{-4}{11}\right)^t \left(\frac{1}{\sqrt{22}}\right) \\ \frac{0}{\sqrt{2}} \end{pmatrix} \tag{3.3.4b}$$

$$\begin{aligned} D_t^2(1,2) &= \sum_{j=1}^{n-1} \lambda_j^{2t} (\langle 1 | \psi_j \rangle - \langle 2 | \psi_j \rangle)^2 \\ &= \left(\frac{7}{11}\right)^{2t} \left(\frac{-4}{2\sqrt{11}}\right)^2 + \left(\frac{-4}{11}\right)^{2t} \left(\frac{6}{\sqrt{22}}\right)^2 = \left(\frac{7}{11}\right)^{2t} \left(\frac{4}{11}\right) + \left(\frac{4}{11}\right)^{2t} \left(\frac{18}{11}\right)^2 \end{aligned} \tag{3.3.5a}$$

$$\begin{aligned} D_t^2(1,3) &= \sum_{j=1}^{n-1} \lambda_j^{2t} (\langle 1 | \psi_j \rangle - \langle 3 | \psi_j \rangle)^2 \\ &= \left(\frac{7}{11}\right)^{2t} \left(\frac{-8}{2\sqrt{11}}\right)^2 + \left(\frac{-4}{11}\right)^{2t} \left(\frac{1}{\sqrt{22}}\right)^2 = \left(\frac{7}{11}\right)^{2t} \left(\frac{16}{11}\right) + \left(\frac{4}{11}\right)^{2t} \left(\frac{1}{22}\right)^2 \end{aligned} \tag{3.3.5b}$$

$$\begin{aligned} D_t^2(2,3) &= \sum_{j=1}^{n-1} \lambda_j^{2t} (\langle 2 | \psi_j \rangle - \langle 3 | \psi_j \rangle)^2 \\ &= \left(\frac{7}{11}\right)^{2t} \left(\frac{-4}{2\sqrt{11}}\right)^2 + \left(\frac{-4}{11}\right)^{2t} \left(\frac{-5}{\sqrt{22}}\right)^2 = \left(\frac{7}{11}\right)^{2t} \left(\frac{4}{11}\right) + \left(\frac{4}{11}\right)^{2t} \left(\frac{25}{22}\right)^2 \end{aligned} \tag{3.3.5c}$$

$$D_t^2(2,4) = D_t^2(2,3), \quad D_t^2(1,4) = D_t^2(1,3), \tag{3.3.5d}$$

$$D_t^2(3,4) = \sum_{j=1}^{n-1} \lambda_j^{2t} (\langle 3 | \psi_j \rangle - \langle 4 | \psi_j \rangle)^2 = 0 \tag{3.3.5e}$$

We see that $d_3$ and $d_4$ are merged even when t = 0. When *t* becomes big, only first term remain significant and:

$$D_t^2(1,3) = D_t^2(1,4) \underset{t\gg 1}{>} D_t^2(2,3) = D_t^2(2,4) \approx D_t^2(2,1) \underset{t\to\infty}{\to} 0 \qquad (3.3.6)$$

We see that $d_2$ forms the center of a "cluster".

This example, thought very simple, may help us to understand what diffusion map is doing. As discussed in Ref. [4-6], since the eigenvalues in Eq. (3.3.2) have the order in Eq.(3.2.4a), the more steps we have, the closer become the related points (data clustering) and the fewer upper components have significant contributions to the distance (dimensional reduction). The whole picture here may be viewed as follows.

1. We define our initial Hilbert state vector as the linear combinations of the eigenvectors with the uniform probabilities:

$$|\Psi^{(0)}\rangle = I\,|\Psi^{(0)}\rangle = \sum_{i=0}^{n} |\psi_i\rangle\langle\psi_i|\Psi^{(0)}\rangle = \sum_{i=1}^{n} \frac{1}{\sqrt{n}} |\psi_i\rangle \qquad (3.3.7)$$

2. Use Markov transition operator to act on it t-times as in Eq. (3.2.17):

$$\hat{P}^t\,|\Psi^{(0)}\rangle = \sum_{i=0} \frac{1}{\sqrt{n}} \hat{P}^t\,|\psi_i\rangle = \sum_{i=0} \frac{1}{\sqrt{n}} \lambda_i^{\,t}\,|\psi_i\rangle \qquad (3.3.8)$$

Because the eigenvalues have the values and order in Eq. (3.2.4a), so when *t* is big enough, only the upper most ones have significant contributions The first term is time-dependent, and eventually will be the only significant term. To better describe the documents locations, we should remove it. If we have many eigenvectors with smaller absolute eigenvalues (smaller than one), we can keep the top few vectors in order to reduce the dimension.

## 3.4. Diffusion Map: a Text Document Example

In this section, we discuss diffusion map in text document space. We will follow the steps in Appendix B of Ref. [6], build transition matrix, find its left/right eigenvectors and make diffusion map.

**Example 3.4.1 The SVD Right Matrix** (see the *Grossman-Frieder example* in Ref [9], Eq. (3.1.2) of Ref. [10], or [12]): The starting point is the document-term matrix *A* (see §3.1, [10]):

$$A^T = \begin{bmatrix} \langle d_1| \\ \langle d_2| \\ \langle d_3| \end{bmatrix} = \begin{bmatrix} 1 & 0 & 1 & 0 & 1 & 1 & 1 & 1 & 1 & 0 & 0 \\ 1 & 1 & 0 & 1 & 0 & 0 & 1 & 1 & 0 & 2 & 1 \\ 1 & 1 & 0 & 0 & 0 & 1 & 1 & 1 & 1 & 0 & 1 \end{bmatrix} \qquad (3.4.1a)$$

Then the right (document-document) matrix *R* is calculated from *A*:

$$R = A^T A = \begin{bmatrix} \langle d_1 | \\ \langle d_2 | \\ \langle d_3 | \end{bmatrix} \left[ | d_1 \rangle, | d_2 \rangle, | d_3 \rangle \right] = \begin{bmatrix} 7 & 3 & 5 \\ 3 & 10 & 5 \\ 5 & 5 & 7 \end{bmatrix} \tag{3.4.1b}$$

We will take this symmetric matrix $R$ as the weight matrix $w$ in Eq. (3.2.1a). We prefer this matrix to the matrix $Q(i, j)$ proposed in Ref. [5]:

$$Q(i, j) = \log\left( \frac{N_{i,j}}{\tilde{N}_i N_j} \right) = where \begin{cases} N_{i,j} : number\ of\ words\ j\ in\ document\ i \\ \tilde{N}_i : total\ number\ of\ words\ in\ document\ i \\ N_j : total\ number\ of\ words\ j\ in\ all\ document \end{cases} \tag{3.4.2}$$

The reasons are:
1. It is not clear how to construct the weight matrix $w$ in Eq. (3.2.1a) from $Q(i, j)$;
2. It is not clear how to deal with the singular cases like $N_{i,j} = 0$.
3. We want to compare diffusion map with SVD, based on document-term matrix $A$.

The sample space of this document space has a basis as in Eq. (3.1.1) with $r = 3$. Its base observable is its document label, valued from 1 to 3:

$$\hat{n} | x) = x | x), \quad P(x | \hat{n} | y) = x\delta_{xy}, \quad \sum_{x=1}^{3} | x)P(x | = 1 \tag{3.4.3}$$

The sample state $P$-ket in this sample space is:

$$| \Omega^{(t)}) = \sum_{x=1}^{3} | x)P(x | \Omega^{(t)}) = \sum_{x=1}^{3} m^{(t)}(x) | x) \tag{3.4.4}$$

Using symmetric matrix $R$ as the weight matrix $w$ in Eq. (3.2.1a), we can build a symmetric matrix $a$ (see Eq. (9) in Appendix B of [6]):

$$a(x, y) = \sqrt{\frac{d(x)}{d(y)}} p(x, y) = \frac{w(x, y)}{\sqrt{d(x)d(y)}} = \frac{R_{xy}}{\sqrt{(\sum_z R_{xz})(\sum_z R_{yz})}} \tag{3.4.5}$$

Here, the values of $d(x)$ can be obtained from Eq. (3.2.1b) and (3.4.1b):

$$d(1) = 15, \quad d(2) = 18, \quad d(3) = 17 \tag{3.4.6a}$$

Using Eq. (3.4.1-3), we get the numeric expression of the symmetric matrix:

$$a = \begin{bmatrix} 7/15 & 3/\sqrt{15*18} & 5/\sqrt{15*17} \\ 3/\sqrt{18*15} & 10/18 & 5/\sqrt{18*17} \\ 5/\sqrt{15*17} & 5/\sqrt{18*17} & 7/17 \end{bmatrix} = \begin{bmatrix} 0.4667 & 0.1826 & 0.3131 \\ 0.1826 & 0.5556 & 0.2858 \\ 0.3131 & 0.2858 & 0.4118 \end{bmatrix} \quad (3.4.6b)$$

From Eq. (3.2.2) and (3.4.1), we get the transition matrix:

$$p = \begin{bmatrix} 7/15 & 1/5 & 1/3 \\ 1/6 & 5/9 & 5/18 \\ 5/17 & 5/17 & 7/17 \end{bmatrix} \quad (3.4.6c)$$

Using online matrix calculator [11], we find the eigenvectors and eigenvalues of $a$:

$$\begin{aligned} \lambda_0 &= 1.0000: \quad | v_0 \rangle = [0.5477, 0.6000, 0.5831]^T \\ \lambda_1 &= 0.3552: \quad | v_1 \rangle = [0.6422, -0.7482, 0.1667]^T \\ \lambda_2 &= 0.0989: \quad | v_2 \rangle = [-0.5363, -0.2832, 0.7951]^T \end{aligned} \quad (3.4.6d)$$

The left and right eigenvectors of $p(x, y)$ are derived from $|v_i\rangle$:

$$\langle \varphi_l | y \rangle = \langle y | v_l \rangle \langle y | v_0 \rangle, \quad \langle x | \psi_l \rangle = \langle x | v_l \rangle \langle x | v_0 \rangle \quad (3.4.7)$$

Using Eq. (3.4.6-7), we get their expressions as follows:

$$\begin{aligned} &\lambda_0 = 1.0000: \\ &\quad | \psi_0 \rangle = [1.0000, 1.0000, 1.0000]^T, \langle \varphi_0 | = [0.3000, 0.3600, 0.3400] \end{aligned} \quad (3.4.8a)$$

$$\begin{aligned} &\lambda_1 = 0.3552: \\ &\quad | \psi_1 \rangle = [1.1725, -1.2470, 0.2859]^T, \langle \varphi_1 | = [0.3517, -0.4489, 0.0927] \end{aligned} \quad (3.4.8b)$$

$$\begin{aligned} &\lambda_2 = 0.0989: \\ &\quad | \psi_2 \rangle = [-0.9791, -0.4720, 1.3635]^T, \langle \varphi_2 | = [-0.2937, -0.1699, 0.4636] \end{aligned} \quad (3.4.8c)$$

One can check they do form an orthonormal vector set (Eq. (13) in Appendix B of [5]):

$$\begin{bmatrix} \langle \varphi_0 | \\ \langle \varphi_1 | \\ \langle \varphi_2 | \end{bmatrix} [| \psi_0 \rangle, | \psi_1 \rangle, | \psi_2 \rangle] = \begin{bmatrix} 1 & 0.00004 & -0.00006 \\ 0 & 0.99994 & 0.00006 \\ 0 & 0.00005 & 0.99987 \end{bmatrix} \approx 1 \quad (3.4.8d)$$

The diffusion distances now can be expressed as:

$$D_t^2(1,2) = \sum_{j=1}^{2} \lambda_j^{2t} (\langle 1 | \psi_j \rangle - \langle 2 | \psi_j \rangle)^2 = (0.3552)^{2t}(5.854) + (0.0989)^{2t}(0.2571) \quad (3.4.9)$$

$$D_t^2(1,3) = \sum_{j=1}^{2} \lambda_j^{2t} (\langle 1 | \psi_j \rangle - \langle 3 | \psi_j \rangle)^2 = (0.3552)^{2t}(0.7861) + (0.0989)^{2t}(5.4878) \quad (3.4.10)$$

$$D_t^2(2,3) = \sum_{j=1}^{2} \lambda_j^{2t} (\langle 2 | \psi_j \rangle - \langle 3 | \psi_j \rangle)^2 = (0.3552)^{2t}(2.3500) + (0.0989)^{2t}(3.3690) \quad (3.4.11)$$

For t > 0, the distance between the three documents are in such an order:

$$D_t^2(1,2) > D_t^2(2,3) > D_t^2(1,3) \quad (3.4.12)$$

In Ref. [10], we have evaluated their distances using a metric tensor based on the same SVD example, and, by using top two eigenvectors, we get the following results:

$$d(1,2) = 1.4547, \;\; d(2,3) = 1.1638, \;\; d(1,3) = 0.5140 \quad (3.4.13a)$$

The documents appear to have the same order of distances in the two methods. This should not be a surprise. In SVD method, the left (term-term) matrix *L* is involved:

$$L = A\,A^T = \left[ | d_1 \rangle, |d_2\rangle, | d_3 \rangle \right] \begin{bmatrix} \langle d_1 | \\ \langle d_2 | \\ \langle d_3 | \end{bmatrix} = \sum_{i=1}^{3} | d_i \rangle \langle d_i | \quad (3.4.13b)$$

The documents were mapped to a new coordinate system derived from the eigenvectors of the left matrix (see [10], (3.2.6a) or (3.2.12); while in diffusion mapping, we use the eigenvectors of the right (document-document) matrix *R*, documents were mapped to a new coordinate system derived from the eigenvectors of the right matrix. Both matrices are based on the same document-term frequency matrix *A* in Eq. (3.4.1), hence they should give similar order of closeness of documents. .

To end this section, we would like to make two comments.

1. The transition operator $\hat{P}$ is NOT an observable of the original document space, since, according to Eq. (3.1.8), it does not commute with the base observable $\hat{X}$ :

$$\hat{P} | x \rangle = \sum_{x',y'=1}^{n} | x' \rangle P_{x'y'} \langle y' | x \rangle = \sum_{x'=1}^{n} | x' \rangle P_{x'x} \quad (3.4.14a)$$

$$\hat{P}\,\hat{X} | x \rangle = x \sum_{x'=1}^{n} | x' \rangle P_{x'x} \neq \hat{X}\,\hat{P} | x \rangle = \sum_{x'=1}^{n} x' | x' \rangle P_{x'x} \Rightarrow \hat{P}\,\hat{X} - \hat{X}\,\hat{P} \neq 0 \quad (3.4.14b)$$

Therefore, *the Hilbert space spanned by the eigenvectors of the transition matrix is NOT equivalent to the sample space of Markov chain*.

2. In Appendix B of Ref. [6], the following *spectral decomposition* is given:

$$P_t(x,y)=\sum_{l=0}^{n-1}\lambda_l^t\psi_l(x)\varphi_l(y) \tag{3.4.15}$$

Or, in bra-ket form in Hilbert space:

$$\hat{P}_t=\sum_{l=0}^{n-1}\lambda_l^t\,|\psi_l\rangle\langle\varphi_l\,| \tag{3.4.16}$$

This *is NOT a transition operator in the sample space*, either. Suppose that the decomposition in Eq. (3.4.15) satisfies the normalization condition Eq. (3.2.3a) for t = 1:

$$\sum_{y=1}^{n}P_1(x,y)=\sum_{l=0}^{n-1}\lambda_l\psi_l(x)\sum_{y=1}^{n}\ \varphi_l(y)=1 \tag{3.4.17}$$

Then, due to Eq. (3.4.2a), it would not satisfy Eq. (3.2.3a) for t > 1:

$$\sum_{y=1}^{n}P_t(x,y)=\sum_{l=0}^{n-1}\lambda_l^t\psi_l(x)\sum_{y=1}^{n}\ \varphi_l(y)<1 \tag{3.4.18}$$

Therefore, the *Hilbert transition operator decomposed in Eq. (3.4.15) is NOT a transition operator in the sample space of Markov chain.*

## 4. Phase Space and Fock Space in Thermophysics

In this section, we discuss some simple examples of many-particle systems in Thermophysics. The particles are identical and not interacting with each other. The system is in equilibrium at a given temperature *T*. We want to find the possible Hilbert state of a single particle using Quantum Mechanics (QM) and the distribution function in sample (or phase) space given by statistic thermodynamics for semi-classical ideal gas. We also want to find the relations between the Fock space of many-particle system in Quantum Field Theories (QFT) and the sample (phase) space in quantum statistics.

### 4.1. The Wave Function of a Particle in Ideal Monatomic Gas

What is the wave function of a single particle of monatomic ideal gas confined in a square well with fixed total number of molecules (*N*) and in equilibrium at temperature *T*? This is a good example of mapping between phase space and Hilbert space. (As we will see in the next section, the system state is in Fock space of many-particle systems).

First, from QM, the Hamiltonian of a single particle is given by:

$$\hat{H} = -\frac{\hbar^2}{2m}\nabla^2 + V(\vec{x}) \tag{4.1.1}$$

The potential is a square well given by:

$$V(\vec{x}) = \begin{cases} \infty \;\; if \;\; x \le 0 \; or \; x \ge a \; or \; y \le 0 \; or \; y \ge b \;\; or \; z \le 0 \; or \; z \ge c \\ \qquad\qquad 0, \quad otherwise \end{cases} \tag{4.1.2}$$

The base wave function (i.e., the eigenvector in coordinate representation; see [14], Eq. (2.2-12)) of the Hamiltonian is:

$$\psi_{\vec{j}}(x) = \langle \vec{x} \mid \vec{j} \rangle = \left(\frac{8}{abc}\right)^{1/2} \sin\left(\frac{j_x \pi}{a} x\right) \sin\left(\frac{j_y \pi}{b} y\right) \sin\left(\frac{j_z \pi}{c} z\right) \tag{4.1.3}$$

The general state ket of the particle is an expansion of base kets:

$$|\psi\rangle = \sum_{\vec{j}} c(\vec{j}) \mid \vec{j} \rangle, \quad \hat{H} \mid \vec{j} \rangle = \varepsilon(\vec{j}) \mid \vec{n} \rangle, \quad \varepsilon(\vec{j}) = \frac{h^2}{8m}\left(\frac{j_x^2}{a^2} + \frac{j_y^2}{b^2} + \frac{j_z^2}{c^2}\right) \tag{4.1.4}$$

How do we find the expansion coefficients? From statistic physics, we know that the ***Boltzmann*** distribution function of monatomic ideal gas can be written as [13]:

$$P(\vec{j} \mid \Omega) = m(\vec{j}) = \frac{\exp[-\varepsilon(\vec{j})/kT]}{Z}, \qquad |\Omega) = \sum_{\vec{j}} |\, \vec{j}) P(\vec{j} \,|\Omega) = \sum_{\vec{n}} m(\vec{j})\, |\vec{j}) \tag{4.1.5}$$

Here, $|\Omega)$ is the sample-ket ($P$-ket) of the sample (phase) space of *a single particle*, and the partition function $Z$ and energy are given by (see [13], §11.5):

$$Z = \sum_{\vec{j}} \exp[-\varepsilon(\vec{j})/kT] = V\left(\frac{2\pi kT}{h^2}\right)^{3/2}, \quad V = abc \tag{4.1.6}$$

Now we can map it to the induced Hilbert space, and get the equivalent state v-ket:

$$|\psi\rangle = \sum_{\vec{j}} c(\vec{j}) \mid \vec{j} \rangle, \quad |\, c(\vec{j}) \,| = \sqrt{P(\vec{j} \mid \Omega)} = \frac{\exp[-\varepsilon(\vec{j})/2kT]}{\sqrt{Z}} \tag{4.1.7}$$

Finally, we have the wave function of each single particle at temperature $T$ as:

$$\psi(\vec{x}) = \langle \vec{x} \mid \psi \rangle = \sum_{\vec{n}} c(\vec{j}) \langle \vec{x} \mid \vec{j} \rangle$$

$$= \sum_{\vec{j}} \frac{\exp[-\varepsilon(\vec{j})/2kT]}{\sqrt{Z}} \left(\frac{8}{abc}\right)^{1/2} \sin\left(\frac{j_x \pi}{a} x\right) \sin\left(\frac{j_y \pi}{b} y\right) \sin\left(\frac{j_z \pi}{c} z\right) \tag{4.1.8}$$

Here we see the power of using abstract *P*-ket of *PBN*. Firstly, to expand the phase space of single particle, Eq. (4.1.5), we do not need to know the expression of base *P*-ket$|\vec{n})$. When we get the mapped Hilbert state, Eq. (4.1.7), it still is an abstract v-ket of *VBN*, which is also representation-independent. Only when we choose to use coordinate representation, we need to know the expression (4.1. 3), that can be derived from QM. Secondly, we now can calculate expectation value of, e.g., the energy, in either space:

$$\langle E \rangle = \langle \psi \,|\, \hat{H} \,|\, \psi \rangle = \sum_{\vec{j}} |\, c(\vec{j})\,|^2 \, \varepsilon(\vec{j}) = P(\Omega \,|\, \hat{E} \,|\, \Omega) = \sum_{\vec{j}} m(\vec{j})\varepsilon(\vec{j}) \tag{4.1.9}$$

## 4.2. Multi-variable Sample Space and Fock Space

In Ref. [10], we proposed that most IR models can be represented using Fock space. For example, a document in a 5-term collection can be expressed as:

$$|\, d\,\rangle \equiv \prod_{\alpha=1}^{5} |\, n_\alpha \rangle_\alpha = |1\rangle_1 \,|1\rangle_2 \,|\,0\rangle_3 \,|\,0\rangle_4 \,|1\rangle_5 \equiv |1,1,0,0,1\rangle \tag{4.2.1}$$

These vectors are eigenvectors of occupation number operators $\hat{\vec{N}} = (\hat{n}_1, \ldots, \hat{n}_t)$:

$$\hat{n}_i \left|\vec{N}\right\rangle = \hat{n}_i \,|\, n_1, n_2, \ldots, n_t \rangle = n_i \,|\, n_1, n_2, \ldots, n_t \rangle \tag{4.2.2a}$$

$$\left|\vec{N}\right\rangle = |\, n_1, n_2, \ldots, n_t \rangle = \prod_{k=1}^{t} |\, n_k \rangle, \quad |\, n_j \rangle \,|\, n_k \rangle = |\, n_k \rangle \,|\, n_j \rangle, \quad [\hat{n}_i, \hat{n}_j] = 0 \tag{4.2.2b}$$

In Ref. [1], we proposed that the base *P*-ket of multiple random variables can be expressed as:

$$|\,\vec{N}) = |\, n_1, n_2, \ldots, n_t) = \prod_{i=1}^{t} |\, n_i)_i, \quad |\, n_i)_i \,|\, n_j)_j = |\, n_j)_j \,|\, n_i)_i \tag{4.2.3a}$$

Theses *P*-kets are eigen-kets of the random observables $\vec{N} = (N_1, N_2, \ldots, N_t)$:

$$N_i \,|\, n_1, \ldots, n_t) = n_i \,|\, n_1, \ldots, n_t), \quad (n_1, \ldots, n_t \,|\, N_i = (n_1, \ldots, n_t \,|\, n_i \tag{4.2.3b}$$

Here we see the equivalence of the two spaces based on shared observables $\hat{\vec{N}}$ and

$$\langle \vec{N}' \,|\, \vec{N} \rangle = \delta_{\vec{N}',\vec{N}}, \sum_{\vec{N}} |\, \vec{N} \rangle \langle \vec{N} \,| = 1 \Leftrightarrow (\vec{N}' \,|\, \vec{N}) = \delta_{\vec{N}',\vec{N}}, \sum_{\vec{N}} |\, \vec{N})(\vec{N} \,| = 1 \tag{4.2.4}$$

Now let us study many-particle systems in Thermophysics. From quantum statistics (see [16], §4 and §5), we know that the grand partition function of a system of many identical particles is defined as:

$$Z_G = \sum_{N,J} \exp[-\beta(E_j - \mu N)] = \sum_{N,J} \langle N, j | \exp[-\beta(\hat{H} - \mu\hat{N})] | N, j\rangle$$
$$= Tr(\exp[-\beta(\hat{H} - \mu\hat{N})] \tag{4.2.5}$$

For any operator $\hat{O}$, the ensample average $\langle\hat{O}\rangle$ is obtained by the representation:

$$\langle\hat{O}\rangle = \frac{Tr\{\exp[-\beta(\hat{H} - \mu\hat{N})]\hat{O}\}}{Tr\{\exp[-\beta(\hat{H} - \mu\hat{N})]\}} \tag{4.2.6}$$

In Fock space, the total Hamiltonian and the operator of total occupation number are:

$$\hat{H} = \sum \hat{n}_j \varepsilon_j, \quad \hat{N} = \sum_j \hat{n}_j, \quad \hat{n}_j | \vec{N}\rangle = \hat{n}_j | n_1, \ldots n_\infty\rangle = n_j | n_1, \ldots n_\infty\rangle \tag{4.2.7}$$

Using Eq. (4.2.2b), we can factor the grand partition function as (see [16], page 37):

$$Z_G = \sum_{\vec{N}} \langle\vec{N} | \exp[-\beta(\hat{H} - \mu\hat{N})] | \vec{N}\rangle = \prod_{i=1}^{\infty} \sum_{n_i} \langle n_i | \exp[-\beta(\varepsilon_i - \mu)n_i] | n_i\rangle$$
$$= \prod_{i=1}^{\infty} Tr\{\exp[-\beta(\varepsilon_i - \mu)\hat{n}_i]\} = \prod_{i=1}^{\infty} Z_i \tag{4.2.8}$$

If an operator is a linear function of occupation numbers in the following form:

$$O(\hat{\vec{N}}) = \sum_{i=1}^{\infty} a_i \hat{n}_i \tag{4.2.9}$$

Then its expectation value can be obtained as:

$$\langle O(\hat{\vec{N}})\rangle = \sum_{i=1}^{\infty} \frac{Tr\{a_i \hat{n}_i \exp[-\beta(\varepsilon_i - \mu)\hat{n}_i]\}}{Z_i} \tag{4.2.10}$$

**Bose-Einstein Distribution**: For Bosons, the occupation numbers are not restricted, so the partition function of single state is given by ([14], [15]:

$$Z_i = Tr\{\exp[-\beta(\varepsilon_i - \mu)\hat{n}_i]\} = \sum_{n=0}^{\infty} \langle n | \exp[-\beta(\varepsilon_i - \mu)\hat{n}] | n\rangle$$
$$= \sum_{n=0}^{\infty} \exp[-\beta(\varepsilon_i - \mu)n] = \left(1 - \exp[\beta(\varepsilon_i - \mu)]\right)^{-1} \tag{4.2.11a}$$

The mean number of occupation number in a single state can be easily obtained as:

$$n(\varepsilon_i) \equiv \langle \hat{n}_i \rangle = \frac{Tr\{\hat{n}_i \exp[-\beta(\varepsilon_i - \mu)\hat{n}_i]\}}{Z_i} = \frac{\sum_{n=0} n \exp[-\beta(\varepsilon_i - \mu)n]}{Z_i}$$
$$= -\frac{\partial}{\beta \partial \mu} \ln Z_i = \frac{1}{\exp[\beta(\varepsilon_i - \mu)] - 1} \qquad (4.2.12a)$$

**Fermi-Dirac Distribution**: For fermions, we can get similar formulas. The only difference is, because no two identical fermions can be in one state, the partition function of a single state now is ([14], [15]):

$$Z_i = Tr\{\exp[-\beta(\varepsilon_i - \mu)\hat{n}_i]\} = \sum_{n=0}^{1} \langle n | \exp[-\beta(\varepsilon_i - \mu)\hat{n}] | n \rangle$$
$$= \sum_{n=0}^{1} \exp[-\beta(\varepsilon_i - \mu)n] = 1 + \exp[-\beta(\varepsilon_i - \mu)] \qquad (4.2.11b)$$

The mean number of occupation number in a single state can be easily obtained as:

$$n(\varepsilon_i) = \frac{\partial}{\beta \partial \mu} \ln Z_i = \frac{1}{\exp[\beta(\varepsilon_i - \mu)] + 1} \qquad (4.2.12b)$$

Using probability bracket notation (*PBN*), we know that the probability of the system at one particle state *j* with energy $\varepsilon_j$ and occupation number $n_j$ is given by (see also [15], §11.6):

$$P(n_j | \Omega_j) = m(n_j) = \frac{\exp[-(n_j \varepsilon_j - \mu n_j)/kT]}{Z_j} \qquad (4.2.13)$$

From Eq. (4.2.13), we can find the expected occupation number of given particle state *j* (see Eq. (4.2.4) and [15], §11.6) for bosons:

$$n(\varepsilon_j) \equiv \langle N_j \rangle = \frac{\sum_{n_j=0}^{\infty} n_j \exp[-(n_j \varepsilon_j - \mu n_j)/kT]}{Z_j} = \frac{1}{\exp[(\varepsilon_j - \mu)/kT] - 1} \qquad (4.2.14)$$

This is consistent with the result comes form Fock space, Eq. (4.2.12a). Same is true for fermions. Similar to previous section, we can map $|\Omega_j)$ to state ket for particles at *j*-th state in Hilbert space:

$$|\psi_j\rangle = \sum_{n_j=0}^{\infty} c(n_j)|n_j\rangle, \quad |c(n_j)| = \sqrt{m(j)} = \sqrt{P(n_j|\Omega_j)} \tag{4.2.15}$$

We can verify that:

$$\begin{aligned}\langle N_j\rangle &= P(\Omega_j \,|\, N_j \,|\, \Omega_j) = \sum_{n_j} P(\Omega_j \,|\, N_j \,|\, n_j) P(n_j \,|\, \Omega_j) \\ &= \sum_{n_j} n_j P(n_j \,|\, \Omega_j) = n(\varepsilon_j)\end{aligned} \tag{4.2.16a}$$

$$\langle \hat{n}_j\rangle = \langle \psi_j \,|\, \hat{n}_j \,|\, \psi_j\rangle = \sum_{n_j,n_j} \langle n_i \,|\, c^*(n_i) n_j\, c(n_j) \,|\, \psi_j\rangle = \sum_{n_j} |c(n_j)|^2\, n_j = n(\varepsilon_j) \tag{4.2.16b}$$

Because the sample (phase) space of the system is the product of the sample (phase) spaces of all single particle states, we can write:

$$\begin{aligned}P(\vec{N}\,|\,\Omega) &= P(n_1, n_2, n_3, \ldots \,|\, \Omega_1, \Omega_2, \Omega_3, \ldots) = \prod_{j=1}^{\infty} P(n_j \,|\, \Omega_j) \\ &= \prod_{j=1}^{\infty} m(n_j) = \frac{\prod_{j=1}^{\infty} \exp[-(n_j\varepsilon_j - \mu n_j)/kT]}{Z_G} = \frac{\exp[-(E(\vec{N}) - \mu N)/kT]}{Z_G}\end{aligned} \tag{4.2.17}$$

Here we have used the eigenvalue of total Hamiltonian:

$$\hat{H}\,|\,\vec{N}\rangle = \sum_{j=1}^{\infty} \hat{n}_j \varepsilon_j \,|\, \vec{N}\rangle = \sum_{j=1}^{\infty} n_j \varepsilon_j \,|\, \vec{N}\rangle = E(\vec{N})\,|\,\vec{N}\rangle \tag{4.2.18}$$

We see that the expectation value of an observable of function $\vec{N}$ can be expressed as:

$$\begin{aligned}\langle O(\hat{\vec{N}})\rangle &= P(\Omega \,|\, O(\hat{\vec{N}}) \,|\, \Omega) = \sum_{\vec{N}} P(\Omega \,|\, O(\hat{\vec{N}}) \,|\, \vec{N}) P(\vec{N} \,|\, \Omega) = \sum_{\vec{N}} P(\Omega \,|\, \vec{N}) O(\vec{N}) P(\vec{N} \,|\, \Omega) \\ &= \sum_{\vec{N}} O(\vec{N}) P(\vec{N} \,|\, \Omega) = \frac{\sum_{\vec{N}} O(\vec{N}) \exp[-(E(\vec{N}) - \mu N)/kT]}{Z_G} \\ &= \frac{\sum_{\vec{N}} \langle \vec{N} \,|\, O(\hat{\vec{N}}) \exp[-\beta(\hat{H} - \mu\hat{N}) \,|\, \vec{N}\rangle}{Z_G} = \frac{Tr\{\hat{O} \exp[-\beta(\hat{H} - \mu\hat{N})\}}{Z_G}\end{aligned} \tag{4.2.19}$$

This means that the expectation value in our *PBN* is consistent with the original expression in Fock space, Eq. (4.6). Now we are ready to find the equilibrium system state at temperature *T* in Fock space:

$$|\Psi\rangle = \sum_{\vec{N}} C(\vec{N})\,|\,\vec{N}\rangle = \sum_{\vec{N}} C(n_1, n_2, \ldots)\,|\,n_1, n_2, \ldots\rangle = \sum_{\vec{N}} \prod_{j=1} c(n_j)\,|\,n_j\rangle \tag{4.2.20}$$

Based on Eq. (4.2.9), we can express the coefficients as:

$$C(\vec{N}) = \sqrt{P(\vec{N}\,|\,\Omega)} = \sqrt{\prod_{j=1}^{\infty} P(n_j\,|\,\Omega)} = \prod_{j=1}^{\infty} \frac{\exp[-(n_j\varepsilon_j - \mu n_j)/2kT]}{\sqrt{Z_j}} \tag{4.2.21}$$

It can be easily shown that:

$$\langle\Psi\,|\,\hat{O}\,|\,\Psi\rangle = \sum_{\vec{N}} O(\vec{N}) C^2(\vec{N}) = \sum_{\vec{N}_j} O(\vec{N}) P(\vec{N}\,|\,\Omega) = P(\Omega\,|\hat{O}\,|\,\Omega) = \langle\hat{O}\rangle \tag{4.2.22}$$

## Summary

In this particle, we used both *PBN* and *VBN* to investigate the relationship between sample space and Hilbert space, and applied it to data clustering and statistic physics.

First, starting from the base observable, we showed how to construct an induced sample space from a Hilbert space, or an induced Hilbert space from a sample space. We also proposed the equivalence between the system states in the two spaces.

Then, using two examples, we discussed Markov chains and diffusion maps in details. The first example was a simple graph, which had a symmetric transition matrix. The second example was from a famous IR example, which used the right matrix *R*, generated from the document-term matrix *A*. The closeness relation of the documents was derived from diffusion map. We saw it was in consistence with the document closeness relation derived from SVD-metric method, which used the left matrix *L*, generated from the same document-term matrix *A*. This implies that we might have two mutually complementary ways of data clustering, based on the same document-term matrix *A*. We also mentioned that the diffusion space (or the Hilbert space spanned by the eigenvectors of transition matrix) is not equivalent to the Markov sample space, because they do not share base observable. Our examples are rather pedagogic than pragmatic. We need to verify our procedure against a fairly large text document collection in order to compare diffusion map with SVD.

Finally, as a *PBN* application to Thermophysics, we derived the wave function of a single particle of semi-classical ideal gas confined in a square well. We also showed the equivalence of the expectation value of operators in Fock space and in our *PBN* for system of identical fermions or bosons.